\documentclass[unsortedadress,preprint,secnumarabic,
nobibnotes,prd]{revtex4}%
\usepackage{amsmath}
\usepackage{amsfonts}
\usepackage{amssymb}
\usepackage{graphicx}%
\begin{document}
\title{Towards SD$p-$brane quantization}
\author{H. Garc\'{\i}a-Compe\'an}
\email{compean@fis.cinvestav.mx}
\affiliation{Departamento de F\'{\i}sica, Centro de Investigaci\'on y de Estudios
Avanzados del IPN,\\
P.O. Box 14-740, 07000 M\'exico D.F., M\'exico}
\author{G. Garc\'{\i}a-Jim\'enez}
\email{ggarcia@fisica.ugto.mx}
\affiliation{Instituto de F\'{\i}sica de la Universidad de Guanajuato,\\
P.O. Box E-143, 37150 Le\'on Gto., M\'exico}
\author{O. Obreg\'on}
\email{octavio@ifug3.ugto.mx}
\affiliation{Instituto de F\'{\i}sica de la Universidad de Guanajuato,\\
P.O. Box E-143, 37150 Le\'on Gto., M\'exico}
\author{C. Ram\'{\i}rez}
\email{cramirez@fcfm.buap.mx}
\affiliation{Facultad de Ciencias F\'{\i}sico Matem\'aticas,\\
Universidad Aut\'{o}noma de Puebla, P.O. Box 1364, 72000 Puebla, M\'{e}xico}
\date{\today}

\begin{abstract}
The quantum mechanical analysis of the canonical hamiltonian description of the
effective action of a SD$p$-brane in bosonic ten dimensional Type II
supergravity in a homogeneous background is given. We find exact solutions for the
corresponding quantum theory by solving the Wheeler-deWitt equation in the late-time
limit of the rolling tachyon. The probability
densities for several values of $p$ are shown and their possible
interpretation is discussed. In the process the effects of electromagnetic fields are also
incorporated and it is shown that in this case the interpretation of tachyon
regarded as ``matter clock'' is modified.

\end{abstract}
\maketitle

\vskip -1truecm

\vskip -1.3truecm \newpage

\setcounter{equation}{0}

\section{Introduction}

In recent years a great deal of attention has been paid to open string tachyon
states, which arise in unstable D$p$-branes or brane-antibrane systems. These
tachyon states have a symmetric potential $V(T)$, with a central maximum and
two symmetric minima, and to it D$(p-1)$ branes are associated, which arise as
a kink interpolating states between these minima. If the boundary conditions
on the tachyon are spacelike, then usual D$(p-1)$-branes arise (for a review,
see \cite{sennonbps}). However if one of these conditions is timelike, then
the tachyon rolls down and time-dependent, spacelike SD$(p-1)$-branes arise
\cite{gs}. These branes are localized in time, i.e. they exist for a short
time and, due to the coupling of the tachyon
with Ramond-Ramond (RR) fields, they carry the same type of charge as D-branes \cite{gs}.
Moreover, the study of the gravitational backreaction of the
tachyon matter has been done.

As soon as the tachyon field rolls down from the top of $V(T)$ towards one of
its minima, it starts to excite open and closed string modes in such a way
that the energy of the unstable D-brane is radiated away. When the tachyon
arrives to its minimum, the radiation is in the form of only closed strings
because open strings cannot exist in the bulk. This has been computed
explicitly, see \cite{malda,rastelli} and references therein. Actually in this
context, a \textit{dual} correspondence between open and closed string modes
has been conjectured, which can be very helpful in the computation of the
effects when tachyon condensates \cite{conjetura}. Such a conjecture states that the
tree level open string theory provides a description of the rolling tachyon system in terms 
of the closed string emission \cite{conjetura}. Moreover this conjecture can be generalized
to include quantum corrections and the full tachyon dynamics \cite{senconj}.

On the other hand, based in previous work \cite{tdynamics}, Sen proposed a field theory 
describing the dynamics of the rolling tachyon \cite{senthree,senfour,senfive}. In this context, he found
that the tachyon field can be interpreted as the {\it time} in quantum cosmology
\cite{sentime}. This was done by coupling the ``tachyon matter'' to a
gravitational field and then performing its canonical quantization. From it, a
Wheeler-deWitt equation turns out, which can be regarded as a time-dependent
Schr\"{o}dinger equation for this gravity-tachyon matter system. The coupling
of the tachyon to gravity has been studied in connection with classical
cosmological evolution \cite{gibb,fquevedo,revsen}. In particular its role
related to inflation has been discussed, see \cite{revgibb} and references therein.

The classical solutions to supergravity including S-branes have been worked out in some
cases, see e.g. \cite{galtsov,martin,fernando}. In Ref. \cite{buchelone}
solutions of the Einstein-Maxwell effective description, in four dimensions,
of the rolling tachyon of the S0-brane proposed in Ref. \cite{gs}, have been found.

Further, in \cite{peettwo,bucheltwo,peetone} the bosonic sector of the
effective ten dimensional supergravity action, coupled to tachyonic matter,
under a maximal symmetric ansatz $ISO(p+1)\times SO(8-p,1)$, has been
considered. There, the time-dependent models were extensively studied and some
classical solutions to supergravity with SD-branes have been worked out. For recent
developments in this direction see Ref. \cite{leblond},

In the present work, we will consider the canonical quantization of the above
mentioned effective action. In quantizing the classical field theory in
\cite{peettwo,bucheltwo,peetone}, we do not expect to describe rigorously
quantum aspects of string theory. Nevertheless, the quantum properties of the
considered field theory seem to be an interesting problem by itself, as
already pointed out by Sen in Ref. \cite{sentime}, where he considers a quantum cosmology
model coupled to the tachyon matter. The SD$p-$brane model
\cite{peettwo,bucheltwo,peetone} we are going to consider can also
be understood as cosmology with dilaton and RR fields, driven by
the tachyon matter. We show that the proposal by Sen, concerning the
interpretation of the tachyon as time, in the late `time' decoupling limit, is
valid for the model under consideration. We find an exact wave function, finite and continuous everywhere for
the corresponding Schr\"{o}dinger equation. The associated probability density shows
an infinity of continuous degenerated maxima describing a path in minisuperspace. Its behavior
with respect to some interesting values of $p$ of the SD$p$-brane is also shown. Moreover,
we will show that even for the next order approximation from the late-time decoupling limit (still
with $T$ large but with nonvanishing
$V(T)$ and $f(T)$, see Ref. \cite{senfour,senfive} and Sec. 2 in Ref. \cite{sentime}), the RR coupling allows an
interpretation of the tachyon as
time, in this case with a Schr\"{o}dinger equation with a time-dependent potential.

It should be remarked however that in the presence of a uniform electric field,
the interpretation of the tachyon as time seems to be spoiled. In the
late-time limit, the tachyon does not decouple from the electric field. This
electric field has been considered, for example in connection with what has
been called the carrollian confinement mechanism for open string states
\cite{yi,revgibb}.

This paper is organized as follows: in section 2 we briefly discuss the model
proposed in Refs. \cite{peettwo,bucheltwo} for the effective action
of a SD$p$-brane. In section 3 we find the hamiltonian constraint for the
SD-brane. Section 4 is devoted to the study of quantum solutions with the
rolling tachyon approximation in the decoupling limit. We also comment about a possible extension
of the interpretation of tachyon as time to a first approximation around the limit at $T\rightarrow\infty$.
In section 5, we
include electric and magnetic fields and exhibit the relevant part of the
Hamiltonian in the late-time limit. Our conclusions are finally presented in
section 6.

\section{The SD$p$-brane Action}

The case we analyze here is that of the low energy effective action of
the closed string interaction with the rolling tachyon matter. This can be
done by means of an action $S_{brane}$ given by the Dirac-Born-Infeld action
of the tachyon plus a Wess-Zumino term describing its coupling to the RR
fields. To this, the action $S_{bulk}$ of the
background ten dimensional supergravity is added, from which we will consider only the
bosonic sector. The action proposed in \cite{peettwo,bucheltwo} for this theory is:%

\begin{align}
S  &  =S_{bulk}+S_{brane},\\
S_{bulk}  &  =\frac{1}{16\pi G_{10}}\int d^{10}x\sqrt{-g}\left(  R-\frac{1}
{2}(\partial\phi)^{2}-\frac{e^{a\phi}}{2(p+2)!}F_{p+2}^{2}\right) ,\\
S_{brane}  &  =\frac{\Lambda}{16\pi G_{10}}\int d^{p+2}x_{\parallel}%
\widehat{\varrho}_{\perp}\left(  -V(T)e^{-\phi}\sqrt{-{\cal A}}\text{ }\right)
+\frac{\Lambda}{16\pi G_{10}}\int\widehat{\varrho}_{\perp}\mathcal{F(}%
T\mathcal{)}dT\wedge C_{p+1},\label{accion}%
\end{align}
where $G_{10}$ is the Newton's constant in the ten-dimensional theory,
$a\equiv(3-p)/2$ is the dilaton coupling, ${\cal A}=\det {\cal A}_{\alpha\beta},$
${\cal A}_{\alpha\beta}=g_{\alpha\beta}e^{\phi/2}+\partial_{\alpha}T\partial_{\beta
}T$ is the tachyon metric, $\mathcal{F(}T\mathcal{)}$ is the factor of
coupling between the tachyon and the RR fields $C_{p+1}$, and $V(T)$ is the
tachyon potential. $\widehat{\varrho}_{\perp}$ is the ``density of branes'',
which does not depend on the parallel coordinates of the brane $x_{\parallel
\text{ }}.$ Greek indices $\alpha,\beta=0,1,\dots,p+1,$ label the time and
parallel coordinates (denoted by $\parallel$) to the SD$p$-brane. Latin
indices $i,j=1,..,8-p$ label the perpendicular coordinates of the brane
(denoted by $\perp$), and capital letters $A,B, \dots,$ etc. stand for
space-time coordinates of the bulk.

Following Refs. \cite{peettwo,bucheltwo}, the simplest model that we can study
is assuming the homogeneous (but non-isotropic) FRW metric. Making the space
decomposition into maximal symmetric direct product $ISO(p+1)\times SO(8-p,1)$
we have the metric%

\begin{equation}
ds^{2}=-N^{2}(t)dt^{2}+a_{\parallel}^{2}(t)dx_{\parallel}^{2}+a_{\perp}%
^{2}(t)dx_{\perp}^{2}, \label{hmetric}%
\end{equation}
where $a_{\parallel}(t)$ and $a_{\perp}(t)$ are the parallel and perpendicular
scaling factors of the brane and $N(t)$ is the lapse function. In
\cite{buchelone,bucheltwo} it was noticed that the SD$p$-brane is not suitable to
be localized by means of a delta function, because it could break at short
scales the $R$-symmetry present in $SO(8-p,1)$. In order to preserve this
symmetry, it was proposed to ``smear out'' the localization of the brane by a
homogeneous distribution along $x_{\perp}.$ Thus the density $\widehat
{\varrho}_{\perp}$ is given by%

\begin{equation}
\widehat{\varrho}_{\perp}=\rho_{\perp}d^{8-p}x_{\perp},
\end{equation}
where $\rho_{\perp}=\rho_{0}\sqrt{g_{H_{8-p}}}=\rho_{0}a_{\perp}^{8-p}$,
$\rho_{0}$ is a constant and $g_{H_{8-p}}$ is the determinant of the metric of
the hyperbolic space perpendicular to the brane. The $\left(  p+2\right)
-$form field strength $F_{p+2}$ is given in terms of the $\left(  p+1\right)
-$form RR potential $C_{p+1}$, which is chosen in a gauge in which the only
nonvanishing component is $C_{12\cdots p+1}=C(t),$%

\begin{equation}
F_{p+2}^{2}=-N^{-2}\dot{C}_{p+1}^{2}=-N^{-2}\dot{C}^{2}.
\end{equation}
In order to preserve homogeneity, the
tachyon field is function only of time $T=T(t)$. Hence the tachyon couples to
RR fields in the following form%
\begin{equation}
dT\wedge C_{p+1}=\dot{T}Cd^{p+2}x_{\parallel}.
\end{equation}
In order to simplify the Lagrangian we can introduce the coordinates
$\beta_{1},\beta_{2}$ defined as,
\begin{equation}
\beta_{1}=\frac{1}{9}\left[  (p+1)\beta_{\parallel}+(8-p)\beta_{\perp}\right],
\end{equation}%
\begin{equation}
\beta_{2}=\beta_{\parallel}-\beta_{\perp},
\end{equation}
where $\beta_{\parallel}=\ln a_{\parallel}$ and $\beta_{\perp}=\ln a_{\perp}.$
Also the space volume is given by $V_{S}=\frac{1}{16\pi G_{10}}\int
d^{p+1}x_{\parallel}d^{8-p}x_{\perp}$, so $S=\int d^{10}x\mathcal{L},$ with
$\mathcal{L=}V_{S}\int dt\ L.$ In these coordinates and with the ansatz
(\ref{hmetric}) we have the Lagrangian
\begin{align}
L  &  =-\frac{e^{9\beta_{1}}}{N}\left[  72\dot{\beta}_{1}^{2}-\frac
{(p+1)(8-p)}{9}\dot{\beta}_{2}^{2}-\frac{1}{2}\dot{\phi}^{2}-\frac{e^{a\phi}%
}{2(p+2)!}\dot{C}^{2}\right]  -\lambda e^{9\beta_{1}-a\phi/2}V(T)\sqrt
{N^{2}e^{\phi/2}-\dot{T}^{2}}\nonumber\\
&  +\lambda e^{(8-p)\left[  \beta_{1}-\frac{1}{9}(p+1)\beta_{2}\right]
}\mathcal{F(}T\mathcal{)}\dot{T}C. \label{lagrangiano}%
\end{align}
In order to manage the square root part of the Lagrangian, we introduce a
Lagrange multiplier $\Omega$ \cite{tseytlin} into the Lagrangian
(\ref{lagrangiano}) as follows,
\begin{align}
L  &  =-\frac{e^{9\beta_{1}}}{N}\left[  72\dot{\beta}_{1}^{2}-\frac
{(p+1)(8-p)}{9}\dot{\beta}_{2}^{2}-\frac{1}{2}\dot{\phi}^{2}-\frac{e^{a\phi}%
}{2(p+2)!}\dot{C}^{2}\right]  -\frac{1}{2}\Omega^{-1}\left(  N^{2}e^{\phi
/2}-\dot{T}^{2}\right) \nonumber\\
&  -\frac{1}{2}\lambda^{2}e^{18\beta_{1}-a\phi}V^{2}(T)\Omega+\lambda
e^{(8-p)\left[  \beta_{1}-\frac{1}{9}(p+1)\beta_{2}\right]  }\mathcal{F}%
(T)\dot{T}C,\label{lagrados}%
\end{align}
where $\lambda=\Lambda\rho_{0}.$ As usual, variating this action with respect
to $\Omega$, ${\frac{\partial L}{\partial\Omega}}=0$, and substituting
$\Omega$ from it into Lagrangian (\ref{lagrados}) the Lagrangian
(\ref{lagrangiano}) follows.

\section{The SD$p$-brane Hamiltonian}

In this section we discuss the canonical hamiltonian formalism of the
Lagrangian (\ref{lagrados}). The resulting hamiltonian constraint will be
used in the next section to give the corresponding Wheeler-deWitt equation.
The canonical momenta obtained from the Lagrangian (\ref{lagrados}) are given by,%
\[
P_{1}=\frac{\partial L}{\partial\dot{\beta}_{1}}=-\frac{144}{N}e^{9\beta_{1}%
}\dot{\beta}_{1},
\]%
\[
P_{2}=\frac{\partial L}{\partial\dot{\beta}_{2}}=\frac{2}{9}\frac
{(p+1)(8-p)}{N}e^{9\beta_{1}}\dot{\beta}_{2},
\]%
\[
P_{\phi}=\frac{\partial L}{\partial\dot{\phi}}=\frac{e^{9\beta_{1}}}{N}%
\dot{\phi},
\]%
\[
P_{C}=\frac{\partial L}{\partial\dot{C}}=\frac{e^{9\beta_{1}+a\phi}}%
{N(p+2)!}\dot{C},
\]
\begin{equation}
P_{T}=\frac{\partial L}{\partial\dot{T}}=\Omega^{-1}\dot{T}+\lambda
e^{(8-p)\left[  \beta_{1}-\frac{1}{9}(p+1)\beta_{2}\right]  }\mathcal{F(}%
T\mathcal{)}C.
\end{equation}

With the constraints $P_{\Omega}=P_{N}=0$ implemented, the Hamiltonian is
given by
\[
H=\dot{\beta}_{1}P_{1}+\dot{\beta}_{2}P_{2}+\dot{\phi}P_{\phi}+\dot{C}%
P_{C}+\dot{T}P_{T}-L
\]
\begin{align}
&  =\frac{N}{2}\left\{  -\frac{1}{144}e^{-9\beta_{1}}P_{1}^{2}+\frac
{9e^{-9\beta_{1}}}{2(p+1)(8-p)}P_{2}^{2}+e^{-9\beta_{1}}P_{\phi}%
^{2}+(p+2)!e^{-(9\beta_{1}+a\phi)}P_{C}^{2}\right\} \nonumber\\
&  +\frac{\lambda^{2}}{2}V^{2}(T)e^{18\beta_{1}-a\phi}\Omega+\frac{N^{2}%
\Omega^{-1}}{2}e^{\phi/2}+\frac{\Omega}{2}\bigg[P_{T}-\lambda e^{(8-p)\left[
\beta_{1}-\frac{1}{9}(p+1)\beta_{2}\right]  }\mathcal{F(}T\mathcal{)}%
C\bigg]^{2}.
\end{align}
After elimination of $\Omega$ by its equation of motion $\partial
H/\partial\Omega=0,$ the Hamiltonian gets the form $H=NH_{0}$, where,%

\begin{align}
H_{0}  &  =-\frac{1}{144}e^{-9\beta_{1}}P_{1}^{2}+\frac{9e^{-9\beta_{1}}%
}{2(p+1)(8-p)}P_{2}^{2}+e^{-9\beta_{1}}P_{\phi}^{2}+(p+2)!e^{-(9\beta
_{1}+a\phi)}P_{C}^{2}\nonumber\\
&  +2e^{\phi/4}\left\{  \lambda^{2}V^{2}(T)e^{18\beta_{1}-a\phi}%
+[P_{T}-\lambda e^{(8-p)\left[  \beta_{1}-\frac{1}{9}(p+1)\beta_{2}\right]
}\mathcal{F(}T\mathcal{)}C]^{2}\right\}  ^{1/2}=0. \label{hamiltonn}%
\end{align}
is the hamiltonian constraint.

It is worth to notice that when this constraint is applied at the quantum level,
the resulting Wheeler-deWitt equation does not provide a time evolution of the
system, and the corresponding wave function is not normalizable. This is known
as the ``time problem'' \cite{kuchar}.

\section{Canonical Quantization}

Exact expressions for the potential $V(T)$ and the coupling factor
$\mathcal{F(}T\mathcal{)}$ are not known. However, their asymptotic form
$V(T)=e^{-\alpha\left\vert T\right\vert /2}$ and $\mathcal{F(}T\mathcal{)=}%
{\rm sign}(T)e^{-\alpha\left\vert T\right\vert /2}$ as $\left\vert T\right\vert
\rightarrow\infty,$ is known from string theory
\cite{senthree,senfour,senfive,sentime}. Thus we
only assume that $V(T)$ has a maximum at $T=0$ and a minimum at $\left\vert
T\right\vert \rightarrow\infty$, where $V(T)=0$. Also, we see that in this
limit the tachyon decouples also from the RR fields as $\mathcal{F}(T)\rightarrow0$.
The canonical hamiltonian (\ref{hamiltonn}) takes in this limit the form%

\begin{equation}
H_{0}=-\frac{1}{144}e^{-9\beta_{1}}P_{1}^{2}+\frac{9}{2}\frac{e^{-9\beta_{1}}%
}{(p+1)(8-p)}P_{2}^{2}+e^{-9\beta_{1}}P_{\phi}^{2}+(p+2)!e^{-\left(
9\beta_{1}+a\phi\right)  }P_{C}^{2}+2eç^{\phi/4}P_{T}=0. \label{hamilzero}%
\end{equation}
The resulting equation is the Wheeler-deWitt equation,
\begin{equation}
\widehat{H}_{0}\Psi=0, \label{qconstra}%
\end{equation}
where $\widehat{H}_{0}$ is given by (\ref{hamilzero}), with $P_{1}%
=-i\frac{\partial}{\partial\beta_{1}},P_{2}=-i\frac{\partial}{\partial
\beta_{2}},P_{C}=-i\frac{\partial}{\partial C}$ and $P_{T}=-i\frac{\partial
}{\partial T}.$ Assuming that the dilaton field is given by its vacuum
expectation value, i.e. $g_{s}=e^{\left\langle \phi\right\rangle }$, where $g_s$ is
the string coupling constant, then $P_{\phi}=0$, and we have (with a
particular factor ordering),
\begin{equation}
e^{-9\beta_{1}}\left[  C_{1}\frac{\partial^{2}\Psi}{\partial\beta_{1}^{2}%
}-C_{2}\frac{\partial^{2}\Psi}{\partial\beta_{2}^{2}}-C_{3}\frac{\partial
^{2}\Psi}{\partial C^{2}}\right]  =iC_{4}\frac{\partial\Psi}{\partial T},
\label{wdw}%
\end{equation}
where $C_{1}=\frac{1}{144},C_{2}=\frac{9}{2(p+1)(8-p)},C_{3}=(p+2)!g_{s}%
^{-a},C_{4}=2g_{s}^{1/4}.$ Now, we see that the Wheeler-deWitt equation
(\ref{qconstra}) leads to a Schr\"{o}dinger-like equation.

Thus in this limit, the tachyon is a scalar field which provides a useful parametrization
of time, because the tachyon momentum enters linearly in (\ref{hamilzero}). This can be interpreted as a
``matter clock''\footnote{The
analysis and criticism of different proposals as time, in particular the
introduction of matter clocks in quantum gravity and quantum cosmology, has
been nicely done in Ref. \cite{kuchar}.}. In string theory, the corresponding
low energy effective action contains the action of the brane, in which
the tachyon arises. This matter accompanies gravitation ($S_{bulk}$) in a
natural and consistent manner. On the other hand, as mentioned, the tachyon momentum appears
linearly in (\ref{wdw}). So it seems that at least some of the criticisms and
problems related to a ``matter clock'' can be
in this case avoided. Moreover, Sen \cite{senfive} showed that for large values of time
$x_{0}$, the classical tachyon solution goes as $T\simeq x_{0}+\mathcal{O}(e^{-\alpha x_{0}%
})$ thus, this result provide us another way to recognize $T$ as a time.

The solution of (\ref{wdw}) is straightforward. Assuming separation of
variables for $\Psi$ is of the form: $\Psi=$ $\psi_{\beta_{1}}(\beta_{1}%
)\psi_{\beta_{2}}(\beta_{2})\psi_{T}(T)\psi_{C}(C)$ we can rewrite the
equation (\ref{wdw}) as
\begin{equation}
e^{-9\beta_{1}}\left[  C_{1}\frac{\psi_{\beta_{1}}^{\prime\prime}}{\psi
_{\beta_{1}}}-C_{2}\frac{\psi_{\beta_{2}}^{\prime\prime}}{\psi_{\beta_{2}}%
}-C_{3}\frac{\psi_{C}^{\prime\prime}}{\psi_{C}}\right]  =iC_{4}\frac{\psi
_{T}^{\prime}}{\psi_{T}}=-\mu,
\end{equation}
where $\mu$ is a separation constant, which we take to be real. Thus, the
tachyon wave function is given by
\begin{equation}
\psi_{T}(T)=e^{i\left(  \mu/C_{4}\right)  T}.
\end{equation}
Similarly, we find for the other field components
\begin{align}
\psi_{\beta_{2}}(\beta_{2})  &  =e^{\pm i\sqrt{\frac{\sigma}{C_{2}}}\beta_{2}%
},\\
\psi_{C}(C)  &  =e^{\pm i\sqrt{\frac{\xi}{C_{3}}}C},
\end{align}
with $\lambda=\xi+\sigma\geq0$. Thus the remaining equation is given by
\begin{equation}
C_{1}\frac{\psi_{\beta_{1}}^{\prime\prime}}{\psi_{\beta_{1}}}+\mu
e^{9\beta_{1}}+\lambda=0.
\end{equation}
This equation has as solution the modified Bessel function
\begin{equation}
\psi_{\beta_{1}}(\beta_{1})=K_{i\nu}\bigg(\frac{8}{3}\sqrt{\mu}e^{\frac{9}{2}
\beta_{1}}\bigg),
\end{equation}
where $\nu=\frac{2}{9}\sqrt{\frac{\lambda}{C_{1}}}.$ The general solutions are
then,
\begin{equation}
\Psi^{\pm}=\mathcal{N}e^{i\left(  \mu/C_{4}\right)  T}e^{\pm i\sqrt{\frac{\xi
}{C_{3}}}C}e^{\pm i\sqrt{\frac{\sigma}{C_{2}}}\beta_{2}}K_{i\nu}\bigg(\frac
{8}{3}\sqrt{\mu}e^{\frac{9}{2}\beta_{1}}\bigg),\label{solgen}%
\end{equation}
where $\mathcal{N}$ is a normalization constant. This is a plane wave, that
represents a free particle, with respect to the variables $C$ and $\beta_{2}$
and with $T$ playing the role of time. In terms of the radii $a_{\parallel}$ and 
$a_{\perp}$ we have,
\begin{equation}
\Psi^{\pm}=\mathcal{N}e^{i(\mu g_{s}^{-1/4})\,T}e^{\pm i\sqrt{\xi/(p+2)!}%
C}\left(  \frac{a_{\parallel}}{a_{\perp}}\right)  ^{\pm\frac{i}{3}%
\sqrt{2(p+1)(8-p)\sigma}}K_{i\nu}\left(  \frac{8}{3}\sqrt{\mu a_{\parallel
}^{p+1}a_{\perp}^{8-p}}\right)  .
\end{equation}
If we compute the expectation value of $a_{\parallel}$, for a certain constant
value of $a_{\perp}$, we have
\begin{equation}
\langle a_{\parallel}\rangle=\mathcal{N}\int_{0}^{\infty}\Phi^{\ast
}a_{\parallel}\Phi da_{\parallel}=\mathcal{N}\int_{0}^{\infty}\left[  K_{i\nu
}\left(  \frac{8}{3}\sqrt{\mu a_{\parallel}^{p+1}a_{\perp}^{8-p}}\right)
\right]  ^{2}a_{\parallel}da_{\parallel}.
\end{equation}
From which we get
\begin{equation}
\langle a_{\parallel}\rangle=\mathcal{N}\sqrt{\pi}\frac{\Gamma\left(  \frac
{2}{p+1}\right)  \Gamma\left(  \frac{2}{p+1}+i\nu\right)  \Gamma\left(
\frac{2}{p+1}-i\nu\right)  }{(3-p)\Gamma\left(  \frac{3-p}{2(p+1)}\right)
}\left(  \frac{9}{64\mu\langle a_{\perp}\rangle^{8-p}}\right)  ^{\frac{2}%
{p+1}}. \label{apar}%
\end{equation}
This relation can also be written as a sort of uncertainty relation between
the two radii $\langle a_{\parallel}\rangle\sim\langle a_{\perp}%
\rangle^{-2\frac{8-p}{p+1}}$, where the proportionality factor is, for $\nu=1$, of the order
of $10^{-2}$ $\mathcal{N}$ and decreases exponentially as $\nu$
increases. Note that the denominator in (\ref{apar}) does not diverge at $p=3$
due to the properties of the Gamma function, in fact $(3-p)\Gamma\left(
\frac{3-p}{2(p+1)}\right)  =2(p+1)\Gamma\left(  \frac{5+p}{2(p+1)}\right)  $.

In Figure 1, we plotted the probability density $|\Psi|^2$ for the physical (`realistic') case of
 $p=3$, where it is shown a continuum of maxima in the
$a_{\perp}-a_{\parallel}$ plane, following a
path in minisuperspace and showing an inverse relation between the two radii given by Eq.
(\ref{apar}). Figure 2 and Figure 3 show two extreme cases with $p=1$ and $p=8$, respectively. These
figures also shown that, for $p=1$, the probability density is almost projected on the
$a_{\perp}$-axis, that is, for large values of $a_{\parallel}$, it is almost independent on it.
Figure 3, is the case for $p=8$ and
it shows that the probability density $|\Psi|^2$ is independent on $a_{\perp}$ and therefore is
projected on the  $a_{\parallel}$-axis. All these figures are plotted for the specific values of
the parameters given by $\mu=0.1$ and $\nu = 0.7$.

\begin{figure}
\begin{center}
\includegraphics[width=10 cm]{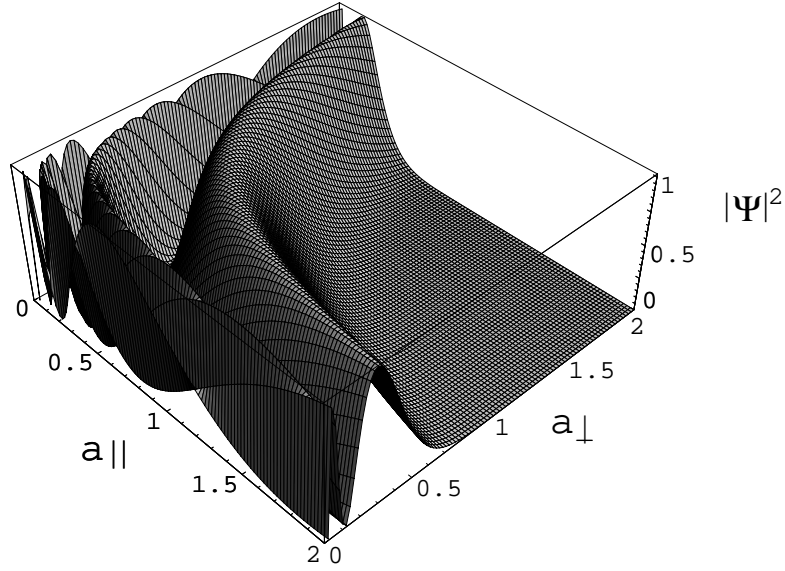}\\
\end{center}
\caption{The figure shows the probability density $|\Psi|^2$ For the 'physical' case of a SD3-brane
($p=3$) (with $\mu = 0.1$ and $\nu =0.7$) and its variation with
respect to the radii $a_{\perp}$ and $a_{\parallel}$. The maxima of the quantum
solution $\Psi$ determines a trajectory in the
$a_{\perp}-a_{\parallel}$ plane.
These maxima satisfy an inverse relation between both radii as shown in
Eq. (27).}
\label{wavef3}
\end{figure}

\begin{figure}
\begin{center}
\includegraphics[width=10 cm]{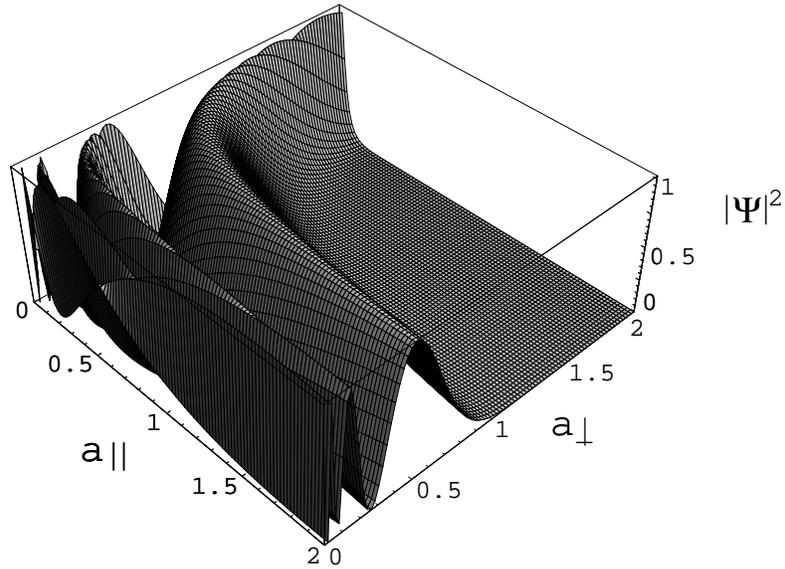}\\
\end{center}
\caption{The probability density $|\Psi|^2$ (also with $\mu = 0.1$ and
$\nu =0.7$) for one extreme case with $p=1$. The solution shows that for this case, the maxima of
$|\Psi|^2$  determines an
evolution which is almost projected on the axis $a_{\perp}$, that is, for large values of
$a_{\perp}$ it is almost independent on $a_{\parallel}$.}
\label{wavef1}
\end{figure}

\begin{figure}
\begin{center}
\includegraphics[width=10 cm]{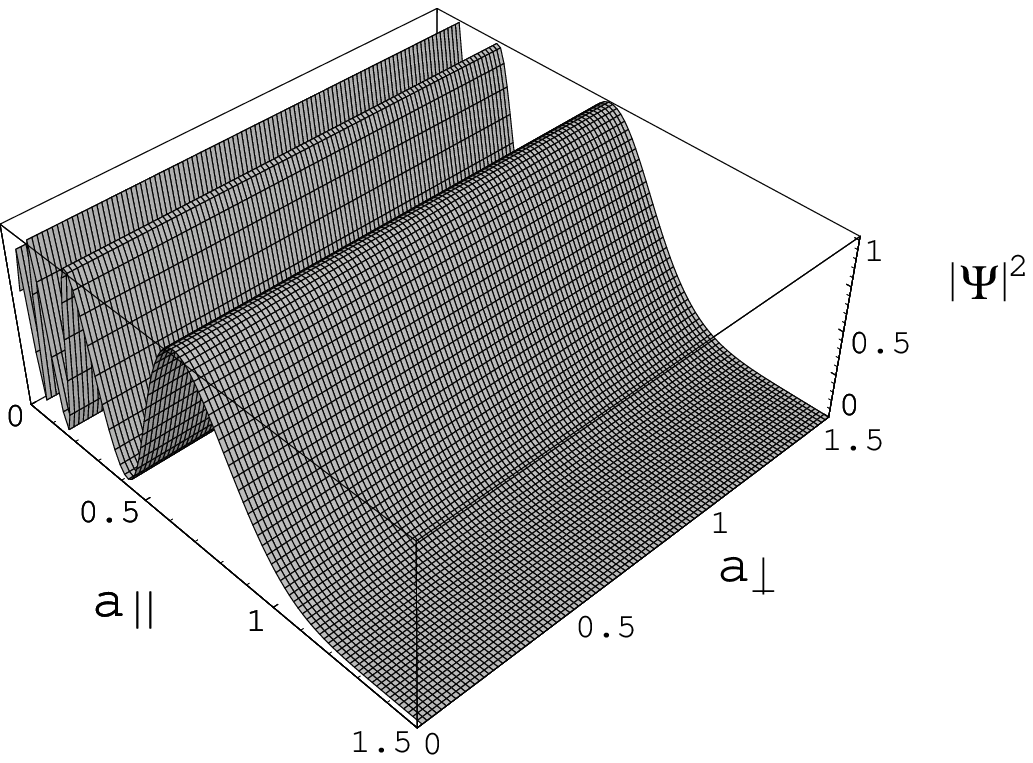}\\
\end{center}
\caption{The figure corresponds with the other extreme case with $p=8$ which shows that the
probability density
$|\Psi|^2$ is independent on  $a_{\perp}$. This corresponds with a wave function $\Psi$
describing an evolution whose maxima are localized around fixed values of $a_{\parallel}$.}
\label{wavef8}
\end{figure}

Let us now consider the next leading order of the approximation of the
hamiltonian (\ref{hamiltonn}), in which $V^{2}(T)$ is neglected with respect
to the $V(T)$ or $\mathcal{F}(T)$. As we mentioned in the introduction, this approximation
corresponds to the first order correction from the late-time decoupling limit with $T$ still large
but nonvanishing $V(T)$ and $f(T)$. This configuration was considered previously by Sen in Refs.
\cite{senfour,senfive,sentime}. In this case we have,
\begin{align}
H_{0}  &  =2e^{\phi/4}P_{T}-\frac{1}{144}e^{-9\beta_{1}}P_{1}^{2}%
+\frac{9e^{-9\beta_{1}}}{2(p+1)(8-p)}P_{2}^{2}+e^{-9\beta_{1}}P_{\phi}%
^{2}+(p+2)!e^{-(9\beta_{1}+a\phi)}P_{C}^{2}\nonumber\\
&  -2\lambda e^{\phi /4} e^{(8-p)\left[  \beta_{1}-\frac{1}{9}(p+1)\beta_{2}\right]
-\frac{\alpha T}{2}}C=0. \label{hamiltonnt}%
\end{align}
After quantization, we obtain from this hamiltonian again a Schr\"{o}dinger
equation, now with a time dependent potential for the RR field. It is interesting to note
that in this case, the term coming from the RR coupling still allows to
interpret the tachyon field as time, because its moment still appears linearly,
but in this case Eq. (\ref{hamiltonnt}) leads to a Schr\"{o}dinger equation with a
time-dependent potential. Of
course in the absence of RR and dilaton fields, the tachyon field is coupled only to gravity
and we recover the situation discussed by Sen in Refs. \cite{senfour,senfive,sentime}.

One way to solve equation (\ref{hamiltonnt}), could be by traying the time-dependent
term as a perturbation.
In this case we could look for a solution of the form,
\begin{equation}
\Psi(T,\beta_1,\beta_2,C)=\psi(T,\beta_1,\beta_2,C)+
e^{-\frac{\alpha T}{2}+\frac{i}{2}\mu g_S^{1/4}T}\psi_1(\beta_1,\beta_2,C).
\label{pert}
\end{equation}
However, when substituted into (\ref{hamiltonnt}), it gives an equation for $\psi_1$
too complicated for an exact solution.
The probability density obtained from (\ref{pert})
$|\Psi|^2\simeq|\psi|^2+2{\rm Re}[\,e^{-\frac{1}{2}(\alpha-i\mu g_S^{1/4})T}\overline\psi\psi_1]$,
contains time dependent interference terms corresponding to interactions of the tachyon matter
(open strings)
with background fields (closed strings). This interference represents a manifestation of the quantum 
backreaction of the tachyon field by the background.

\section{Inclusion of electromagnetic fields}

We want to see in this section how the tachyon dynamics is modified in the
presence of electromagnetic fields. Let us consider the case in which electric
and magnetic fields $f_{\alpha\beta}$ are included. The brane action $S_{Brane}$ from Eq.
(\ref{accion}) is modified as follows \cite{senmuk,roy,rey},

\begin{equation}
S_{Brane}=\frac{\Lambda}{16\pi G_{10}}\int d^{p+2}x_{\parallel}\widehat
{\varrho}_{\perp}\left(  -V(T)e^{-\phi}\sqrt{-{\cal A}}\text{ }\right)
+\frac{\Lambda}{16\pi G_{10}}\int\widehat{\varrho}_{\perp}\mathcal{F(}%
T\mathcal{)}dT\wedge C_{p+1}\wedge e^{f},\label{accionb}%
\end{equation}
where now the tachyon metric is ${\cal A}_{\alpha\beta}=g_{\alpha\beta}e^{\phi
/2}+\partial_{\alpha}T\partial_{\beta}T+f_{\alpha\beta}$ and $f=f_{\alpha
\beta}dx^{\alpha}\wedge dx^{\beta}$. For simplicity, we will consider only one
nonvanishing component for the electric and magnetic fields, $E=f_{01}%
=\partial_{0}A_{1}$ and $B=f_{12}=-\partial_{2}A_{1}$. With this choice, the
exponential $e^{f}$ in the last term of action (\ref{accionb}) contributes
only with a factor one. This can be obtained by direct calculation or
following \cite{peetone}, taking into account the ansatz $ISO(p+1)\times
SO(8-p,1)$. After integration of the space coordinates, we get the
Lagrangian
\begin{align}
&  L_{brane}=\lambda e^{(8-p)\left[  \beta_{1}-\frac{1}{9}(p+1)\beta
_{2}\right]  }\mathcal{F}(T)\dot{T}C\nonumber\\
&  -\lambda\,e^{9\beta_{1}-a\phi/2}\,V(T)\left[  \bigg(N^{2}e^{\phi/2}-\dot{T}%
^{2}\bigg)\bigg(1+e^{-2\left(  \beta_{1}+\frac{8-p}{9}\beta_{2}+\phi/4\right)  }%
B^{2}\bigg)-E^{2}\right]  ^{1/2}.
\end{align}
Making the same procedure of introducing a Lagrange multiplier $\Omega$ we
found in the late-time limit ($V(T)\rightarrow0$ as $\left\vert T\right\vert
\rightarrow\infty$) that the relevant part of the hamiltonian turns out to be,%

\begin{equation}
H_{\substack{brane\\\left\vert T\right\vert \rightarrow\infty}}=2e^{\phi
/4}\left[  \left(  1+e^{-2\left(  \beta_{1}+\frac{8-p}{9}\beta_{2}%
+\phi/4\right)  }B^{2}\right)  P_{T}^{2}+\Pi^{2}\right]  ^{1/2},\label{spoil}%
\end{equation}
where $\Pi$ is the momentum conjugated to $A_{1}$. From this expression, we
see that the tachyon would decouple only if $\Pi$ vanishes. Thus, under the
presence of electromagnetic fields, the tachyon cannot be identified with time
in the sense of a Schr\"{o}dinger-type equation even in the late-time limit.


\section{Conclusions}

In this work, we have provided an exact solution to the canonical quantization
of the SD$p$-brane model \cite{gs,buchelone,peettwo,bucheltwo,peetone}. For this effective action,
a Wheeler-deWitt equation has been obtained from the
hamiltonian analysis. Following Ref. \cite{tseytlin}, the square root in the
tachyonic matter action (\ref{lagrangiano}) was eliminated by the introduction
of a Lagrange multiplier $\Omega$. From the resulting action the Hamiltonian
(\ref{hamiltonn}) has been computed and the decoupling late-time limit
($\left\vert T\right\vert \rightarrow\infty$) has been done. Even though we
have considered the canonical quantization of the effective action with a
maximally symmetric metric (\ref{hmetric}), the quantum version of this field
theory and in particular of the model under consideration is interesting on
its own right \cite{sentime}. Moreover, it could provide some insight on
string theory beyond the classical limit.

Further we show that the proposal by Sen, concerning the interpretation of the
tachyon as time, in the late-time decoupling limit, is valid for this model.
In this limit we find an exact wave function for the corresponding Schr\"{o}dinger equation.
The associated probability density is a finite and continuous function of the radii
$a_{\parallel}$ and $a_{\perp}$, it shows (a non-singular) continuum of maxima along a
definite trajectory, in such a way that if the mean value of one of the radii
increases, the mean value of the other one decreases, as shown in Figure 1 for
$p=3$ and in Figure 2 and Figure 3 for the extreme cases of $p=1$ and $p=8$, respectively. We have
also considered the situation beyond the late-time decoupling limit in which still $T$ is large but
$V(T)$ and $f(T)$ are nonvanishing. The coupling of the tachyon with the RR fields allows us still to
interpret the tachyon as time. However in this case the Wheeler-deWitt equation (\ref{hamiltonnt})
leads to a Schr\"{o}dinger equation with a
time-dependent potential. This situation has been already discussed in Refs.
\cite{senfour,senfive,sentime} at the classical level. If quantum corrections of the
string theory have to be taken into account and if the open-closed duality holds (see remarks of review, \cite{revsen}),
it would be very interesting to explore if solutions of the Schr\"{o}dinger equation  (\ref{hamiltonnt}), or its generalizations (representing open-closed states), correspond to a description (at the lowest level) of the physics of the quantum string theory associated to SD$p$-branes.

Finally, we have also shown that in the presence of electromagnetic fields, the
interpretation of the tachyon as time seems to be spoiled. Indeed, as can be
seen from Eq. (\ref{spoil}) that even in the late-time limit the tachyon does
not decouple from the electric field.

\vskip2truecm
\centerline{\bf Acknowledgments} This work was supported in part by CONACyT
M\'{e}xico Grants Nos. 37851E and 41993F, PROMEP and Gto. University Projects.


\vskip2truecm



\begin{references}
\bibitem{sennonbps} A. Sen, ``Non-BPS States and Branes in String Theory",
Published in Cargese 1999, {\it Progress in string theory and M-theory}, pp
187-234, hep-th/9904207.
\bibitem{gs} M. Gutperle and A. Strominger, ``Spacelike Branes'' JHEP {\bf 0204} (2002)
018, hep-th/0202210.
\bibitem{malda} N. Lambert, H. Liu and J. Maldacena, hep-th/0303139.
\bibitem{rastelli} D. Gaiotto, N. Itzhaki and L. Rastelli, Nucl. Phys. B {\bf 688}
(2004) 70, hep-th/0304192.
\bibitem{conjetura} A. Sen, Phys. Rev. D {\bf 68} (2003) 106003,
hep-th/0305011; Phys. Rev. Lett. {\bf 91} (2003) 181601, hep-th/0306137.
\bibitem{senconj} A. Sen, ``Open Closed Duality: Lessons from Matrix Model'',  Mod. Phys. Lett. A {\bf 19} (2004) 841, hep-th/0308068. 
\bibitem{tdynamics} M.R. Garousi, Nucl. Phys. B {\bf 584} (2000) 284; E.A. Bergshoeff, M. de Roo, T.C. de Wit,
E. Eyras and S. Panda, JHEP {\bf 0005} (2000) 009; J. Kluson, Phys. Rev. D {\bf 62} (2000) 126003; G.W. Gibbons,
K. Hori and P. Yi, Nucl. Phys. B {\bf 596} (2001) 136.
\bibitem{senthree} A. Sen, ``Rolling Tachyon'', JHEP {\bf 0204} (2002) 048,
hep-th/0203211.
\bibitem{senfour} A. Sen, ``Tachyon Matter'', JHEP {\bf 0207} (2002) 065,
hep-th/0203265.
\bibitem{senfive} A. Sen, ``Field Theory of Tachyon Matter'', Mod. Phys. Lett. A
{\bf 17} (2002) 1797, hep-th/0204143.
\bibitem{sentime} A. Sen, ``Time and Tachyon'', Int. J. Mod. Phys. A {\bf 18}
(2003) 4869, hep-th/0209122.
\bibitem{fquevedo} F. Quevedo, ``Lectures on String/Brane Cosmology'', Class.
Quant. Grav. {\bf 19} (2002) 5721, hep-th/0210292.
\bibitem{revsen} A. Sen, ``Remarks on Tachyon Driven Cosmology'', Talk at Nobel
Symposium on Cosmology and String Theory and IIT Kanpur workshop on String Theory,
hep-th/0312153.
\bibitem{gibb} G.W. Gibbons, ``Cosmological Evolution of the Rolling Tachyon'',
Phys. Lett. B {\bf 537} (2002) 1, hep-th/0204008.
\bibitem{revgibb} G.W. Gibbons, ``Thoughts of Tachyon Cosmology'', Class. Quant.
Grav. {\bf 20} (2003) S321, hep-th/0301117.
\bibitem{galtsov} C.-M. Chen, D.V. Galtsov and M. Gutperle, ``S-branes Solutions
in Supergravity Theories´´, Phys. Rev. D {\bf 66} (2002) 024043, hep-th/0204071.
\bibitem{martin} M. Kruczenski, R.C. Myers and A.W. Peet, JHEP {\bf 0205} (2002) 039, 
hep-th/0204144.
\bibitem{fernando} F. Quevedo, G. Tasinato and I. Zavala C., ``$S$-branes,
Negative Tension Branes and Cosmology'', hep-th/0211031.
\bibitem{buchelone} A. Buchel, P. Langfelder and J. Walcher, ``Does the Tachyon
Matter?'', Annals Phys. {\bf 302} (2002) 78, hep-th/0207235; A. Buchel and J.
Walcher, ``The Tachyon does Matter'', Fortsch. Phys. {\bf 51} (2003) 885,
hep-th/0212150.
\bibitem{peettwo} F. Leblond and A.W. Peet, ``SD-brane Gravity Fields and Rolling
Tachyons'', JHEP {\bf 0304} (2003) 048, hep-th/0303035.
\bibitem{bucheltwo} A. Buchel and J. Walcher, ``Comments on Supergravity
Description of S-branes'', JHEP {\bf 0305} (2003) 069, hep-th/0305055.
\bibitem{peetone} F. Leblond and A.W. Peet, ``A Note on the Singularity Theorem
for Supergravity SD-branes'', JHEP {\bf 0404} (2004) 022, hep-th/0305059.
\bibitem{leblond} F. Leblond, ``Mirage Resolution of Cosmological Singularities'',
hep-th/0403221.
\bibitem{yi} G.W. Gibbons, K. Hashimoto and P. Yi, ``Tachyon Condensates,
Carrollian Contraction of Lorentz Group, and Fundamental Strings'', JHEP {\bf
0209} (2002) 061, hep-th/0209034.
\bibitem{tseytlin} A.A. Tseytlin, Nucl. Phys. B {\bf 469} (1996) 51.
\bibitem{kuchar} K.V. Kuchar, ``Time and Interpretations in Quantum Gravity'', in
proceedings of the 4th conference on General Relativity and Relativistic
Astrophysics, eds G. Kunstatter, D. Vincent and J. Williams (World Scientific,
Singapore 1992).
\bibitem{senmuk} P. Mukhopadhyay and A. Sen, JHEP {\bf 0211} (2002) 047,
hep-th/0208142.
\bibitem{roy} S. Bhattacharya, S. Mukherji and S. Roy, ``On the Effective Action
of Space-like Brane'', Phys. Lett. B {\bf 584} (2004) 163, hep-th/0308069.
\bibitem{rey} S.-J. Rey and S. Sugimoto, ``Rolling Tachyon with Electric and
Magnetic Fields- T-duality Approach-'', Phys. Rev. D {\bf 67} (2003) 086008,
hep-th/0301049.
\end{references}
\end{document}